\documentclass[11pt,a4paper]{article}
\usepackage{jcappub}
\usepackage{bm}
\usepackage{amsmath}
\usepackage{amssymb}
\usepackage{amsbsy}
\usepackage{subfigure}
\usepackage{afterpage}
\usepackage{hyperref}
\usepackage[multiple]{footmisc}

\renewcommand\({\left(}
\renewcommand\){\right)}
\renewcommand\[{\left[}
\renewcommand\]{\right]}

\newcommand{\bp}{{\bf p}}
\newcommand{\bx}{{\bf x}}

\newcommand{\bv}{{\bf v}}
\newcommand{\bk}{{\bf k}}
\newcommand{\bq}{{\bf q}}
\newcommand{\be}{{\bf e}}
\newcommand{\bB}{{\bf B}}
\newcommand{\bP}{{\bf P}}
\newcommand{\bD}{{\bf\Delta}}

\newcommand{\sH}{{\sf H}}
\newcommand{\sL}{{\sf L}}
\newcommand{\sM}{{\sf M}}
\newcommand{\sV}{{\sf V}}
\newcommand{\sD}{{\sf D}}

\newcommand{\half}{{\textstyle\frac{1}{2}}}

\newcommand{\leftpartial}{\overleftarrow{\partial}}
\newcommand{\rightpartial}{\overrightarrow{\partial}}

\newcommand{\exclude}[1]{}

\definecolor{gre}{rgb}{0,0.4,0.3}

\begin{document}
\subheader{\hfill MPP-2018-14}


\title{Liouville term for neutrinos:
  Flavor structure and wave interpretation}

\author[a]{Tobias Stirner,}
\author[b]{G\"unter Sigl}
\author[a]{and Georg Raffelt}

\affiliation[a]{Max-Planck-Institut f\"ur Physik (Werner-Heisenberg-Institut),\\
F\"ohringer Ring 6, 80805 M\"unchen, Germany}

\affiliation[b]{II.~Institut f\"ur Theoretische Physik, Universit\"at Hamburg,\\
Luruper Chaussee~149, 22761~Hamburg, Germany}

\emailAdd{stirner@mpp.mpg.de}
\emailAdd{guenter.sigl@desy.de}
\emailAdd{raffelt@mpp.mpg.de}

\abstract{Neutrino production, absorption, transport, and flavor
  evolution in astrophysical environments is described by a kinetic
  equation $D\varrho=-i[\sH,\varrho]+{\cal C}[\varrho]$.  Its basic
  elements are generalized occupation numbers $\varrho$, matrices in
  flavor space, that depend on time $t$, space~$\bx$, and momentum
  $\bp$. The commutator expression encodes flavor conversion in terms
  of a matrix $\sH$ of oscillation frequencies, whereas
  ${\cal C}[\varrho]$ represents source and sink terms as well as
  collisions.  The Liouville operator on the left hand side involves linear
  derivatives in $t$, $\bx$ and $\bp$. The simplified expression
  $D=\partial_t+\hat\bp\cdot{\partial}_\bx$ for ultra-relativistic
  neutrinos was recently questioned in that flavor-dependent
  velocities should appear instead of the unit vector
  $\hat\bp$. Moreover, a new damping term was postulated as a result.
  We here derive the full flavor-dependent velocity structure of the
  Liouville term although it appears to cause only higher-order
  corrections. Moreover, we argue that on the scale of the neutrino
  oscillation length, the kinetic equation can be seen as a first-order
  wave equation.}

\maketitle


\section{Introduction}

Neutrino flavor oscillations \cite{Gribov:1968kq} are one example for
the propagation of a multi-component wave
$\psi=(\psi_1,\ldots,\psi_N)$, where the $N$ components obey different
dispersion relations.  Other examples include the polarization
components of electromagnetic waves, notably the Faraday effect, or
more speculative oscillations between photons and hidden photons
\cite{Redondo:2015iea} or between photons and axion-like
particles~\cite{Raffelt:1987im}. Typically one considers the evolution
of polarization or flavor from a source along a trajectory to a
detector. The dispersion relations usually depend on the frequency of
the radiation so that the expected result depends on the source
spectrum and the detector energy resolution. If $s$ is a coordinate
along the beam, for every frequency $\omega$ one needs to solve an
equation of the form\footnote{In the context of neutrino flavor
  oscillations, the coordinate $s$ is usually interpreted as ``time of
  propagation'' along the beam and the equation is interpreted as a
  Schr\"odinger equation for the flavor content of a given neutrino,
  whereas in the context of the astrophysical Faraday effect one
  always interprets $s$ as a distance along the beam.  The physical
  result is of course the same.}\footnote{We use units in which the
  vacuum speed of light and Planck's constant are unity, $c=\hbar=1$.}
\begin{equation}\label{eq:eom0}
  i\partial_s\psi=\sH\psi\,,
\end{equation}
where $\sH$ is a Hermitean $N{\times}N$ matrix. In the propagation
basis, where $\sH$ is diagonal, its entries are the wave numbers of
the $N$ branches of the dispersion relation with frequency $\omega$.
Solving equation~\eqref{eq:eom0} can be
complicated when $\sH$ varies as a function
of $s$. In particular, if different branches of the dispersion relation cross
(``avoided level crossing'') one can get complete flavor
conversion\footnote{The terminology of ``flavor oscillations'' is actually a bit of
  a misnomer in this context~\cite{Smirnov:2016xzf}.}
even for a small mixing angle---the
celebrated MSW effect~\cite{Wolfenstein:1977ue,Mikheev:1986gs}.
It can also arise for photons in the astrophysical Faraday
context~\cite{Broderick:2009ix, Dasgupta:2010ck}.

A more sophisticated approach is needed for a class of problems where
neutrinos scatter many times after production. One generic example is
sterile-neutrino production in the early universe by oscillations and
collisions \cite{Dolgov:1980cq,Barbieri:1990vx}
by what has come to be called the Dodelson-Widrow
mechanism \cite{Dodelson:1993je}.
On the most elementary level we may
think of a single-particle neutrino state with a flavor content
initially described by the amplitudes $\psi$. Subsequent
flavor-dependent collisions decohere the flavor content as it gets
entangled with the environment and we need to switch to a
density-matrix description in the form $\rho_{ij}=\psi_j^*\psi_i$ with
$i,j=1,\ldots,N$.  The oscillation part of the evolution corresponding
to equation~\eqref{eq:eom0} is the usual commutator expression
$\partial_t\rho=-i[\sH,\rho]$ that applies both to pure and mixed
states.

The decoherence part of the evolution has been formulated in different
ways \cite{Manohar:1986gj,Stodolsky:1986dx,Thomson:1991xq}.  The key
point is that in the interaction basis, the off-diagonal elements of
$\rho$ are damped if the scattering amplitudes for the different
flavors are different. A compact way to express this behavior is in
terms of Lindblad operators ~\cite{Lindblad:1975ef,Alicki:2007}, so
overall we have~\cite{Benatti:2000ph,Ohlsson:2000mj}
\begin{equation}\label{eq:eom1}
  \partial_t\rho=-i[\sH,\rho]-[\sL,[\sL,\rho]]\,,
\end{equation}
where $\sL$ is a Hermitean $N{\times}N$ matrix. Notice that the
commutator structure preserves ${\rm tr}\,\rho=1$, i.e., our single
neutrino is not absorbed in this model, only flavor coherence is
damped.\footnote{In reference~\cite{Ohlsson:2000mj} an ensemble of
  neutrinos was considered with a Gaussian distribution of oscillation
  frequencies. Even without a Lindblad term, the average $\rho$ matrix
  loses its off-diagonal terms because of dephasing of different
  neutrinos and it was argued that this effect was equivalent to the
  loss of coherence by a Lindblad term.  However, apparent decoherence
  caused by dephasing is reversible, for example by a detector with
  sufficient energy resolution, and also does not increase the von
  Neumann entropy of the ensemble. Therefore, one should always
  carefully distinguish between ``kinematical decoherence'' caused by
  dephasing of many neutrinos or different Fourier components of a
  single-neutrino wavepacket, and ``dynamical decoherence'' caused by
  irreversible entanglement with the environment.}  Moreover, the von
Neumann entropy $-{\rm tr}\,(\rho\log\rho)$ increases monotonically
thanks to $\sL$ being self-adjoint. In the context of active-sterile
oscillations, $\sH$ is diagonal in the propagation basis whereas $\sL$
is diagonal in the interaction basis, so $\sH$ and $\sL$ do not
commute. The final asymptotic $\rho$ matrix which commutes with both
$\sH$ and $\sL$ would be proportional to the unit matrix, representing
flavor equilibrium.

The damping of flavor coherence arises more directly if one
considers a kinetic equation for the neutrino distribution. It is
standard to use occupation numbers $f_{\bp}$ to describe the evolution
of a quantum field in terms of a Boltzmann kinetic equation. In a
seminal paper, Dolgov~\cite{Dolgov:1980cq} extended this description
to mixed neutrinos in terms of matrices $\varrho_{\bp}$ which
are generalized occupation numbers.\footnote{We use the symbol $\varrho$ for
  generalized occupation numbers (``matrix of densities'')
  in contrast to $\rho$ as in
  equation~\eqref{eq:eom1} which is a single-particle density matrix.
  In practice, they differ by normalization with
  ${\rm tr}\,\rho=1$, whereas $0\leq{\rm tr}\,\varrho_\bp\leq N$.
  Moreover, ${\rm tr}\,\varrho_\bp$ changes under the kinetic
  evolution by source and sink terms.}
It is assumed that the different $\bp$ components of the
quantum field decohere quickly (``molecular chaos''), i.e.,
correlations between different $\bp$ modes are ignored, whereas
flavor coherence survives on the relevant time scale and is followed
explicitly in terms of the off-diagonal $\varrho_\bp$-components. For
neutrinos in the early universe, Dolgov's
equation is~\cite{Dolgov:1980cq,Barbieri:1990vx}
\begin{equation}\label{eq:eom2}
  \partial_t\varrho_\bp-H\,|\bp|\,\partial_{|\bp|}\varrho_\bp
  =-i[\sH_\bp,\varrho_\bp]+{\cal C}[\varrho_{\bp'},\bar\varrho_{\bp'}]\,,
\end{equation}
where $H$ is the Hubble expansion parameter and we have
assumed isotropy. The crucial new ingredient is the collision term,
where scattering on electrons and positrons as well as pair
processes $e^+e^-\leftrightarrow \nu\bar\nu$ were explicitly included.
There is an equation of type~\eqref{eq:eom2} for every neutrino mode
$\varrho_\bp$ and anti-neutrino mode $\bar\varrho_\bp$.
They are all coupled by the collision term. Key for the approach to
equilibrium are scattering amplitudes that distinguish between
different flavors and thus introduce a nontrivial flavor structure of
the collision term~\cite{Dolgov:1980cq,Barbieri:1990vx}.

These and similar discussions~\cite{McKellar:1992ja} explicitly ignore
degeneracy effects and can be seen as describing single-neutrino
states. However, in compact astrophysical objects such as
core-collapse supernovae or neutron-star mergers, neutrinos can be
degenerate. By considering flavor mixing among quantum fields instead
of wave functions one finds that the oscillation term in
equation~\eqref{eq:eom2} remains unchanged, in particular when ${\rm
  tr}\,\varrho_\bp>1$, i.e., a given mode is occupied by more than one
neutrino~\cite{Raffelt:1991ck}.  Likewise, the formal appearance of
the collision term remains the same, yet each process such as elastic
scattering and pair or $\beta$ processes now include initial-state
occupation numbers and/or final-state Pauli-blocking factors in terms
of $\varrho_\bp$ matrices. Therefore, the collision term shows a
nontrivial flavor structure beyond flavor-dependent scattering
amplitudes
\cite{Raffelt:1992uj,Sigl:1992fn,Rudzsky:1990,Yamada:2000za,Vlasenko:2013fja}.

In a toy model of a simplified collision term which does not couple
different momentum modes, one can show explicitly the emergence of the
Lindblad structure of equation~\eqref{eq:eom1} in the form of a double
commutator for the damping of flavor coherence \cite{Raffelt:1992uj}.
In general, of course, flavor coherence is not damped separately for
each $\varrho_\bp$ and actually can temporarily increase for some
range of modes. It is noteworthy, however, that the kinetic equation
reproduces the expected evolution of the appropriate thermodynamic
potential, assuming the collision term arises from the interaction
with a thermal background medium that can exchange energy and lepton
number with the neutrino gas \cite{Sigl:1992fn}.  In other words,
flavor decoherence of the ensemble follows from the kinetic equation
without further ado, a point that does not seem to be controversial in
the literature.

Thus far we have focussed on flavor conversion driven by neutrino
masses. However, neutrino masses unavoidably couple positive with
negative helicity states, for example implying small neutrino
electromagnetic dipole and transition moments \cite{Giunti:2014ixa}.
Moreover, if the
background medium is not isotropic it induces transitions between
helicity states which, for Majorana neutrinos, effectively implies
transitions between neutrinos and antineutrinos.  Such effects can be
included by considering larger $\varrho_\bp$ matrices that may include
sterile neutrinos (Dirac case) or neutrinos and antineutrinos in a
single $\varrho_\bp$ matrix (Majorana case) \cite{Vlasenko:2013fja,
  Giunti:2014ixa, Lim:1987tk, Studenikin:2004bu, Volpe:2013jgr,
  Serreau:2014cfa, Kartavtsev:2015eva, Dobrynina:2016rwy}.  On the
other hand, if we take the background medium to be isotropic, angular
momentum conservation precludes such effects which anyway always seem
to be negligibly small in practice. For simplicity we will thus ignore
neutrino helicity conversion.

Somewhat surprisingly, it is not the collision or oscillation terms,
but the Liouville term that has aroused some doubts in the recent
literature \cite{Hansen:2016klk}.  If we consider unmixed particles
such as electrons, and if we include weak inhomogeneities, one uses
space-dependent occupation numbers $f_{t,\bx,\bp}$. This construction
makes physical sense when the spatial variation is on scales much
larger than a typical radiation wavelength so that the uncertainty
relation between the non-commuting variables $\bx$ and $\bp$ does not
impose a serious limitation.  Using such Wigner functions allows one
to calculate averages in terms of classical phase-space integrations
instead of invoking rigorous quantum-mechanical expectation values
\cite{Wigner:1932eb,Hillery:1983ms,Lee:1995}.

The kinetic equation is then of the form\footnote{We use
  $\partial_\bx$ to denote the gradient with regard to $\bx$.  It is
  identical with the vector ${\bm\nabla}_\bx$.}  $\partial_t
f_{t,\bx,\bp}+\bv\cdot{\partial}_\bx f_{t,\bx,\bp}
-{\dot\bp}\cdot{\partial}_\bp f_{t,\bx,\bp}={\cal C}[f]$. The last
term on the left hand side (lhs) represents momentum changes by coherent external
forces that are not part of the microscopic collision term. This
includes cosmic expansion as in Dolgov's equation~\eqref{eq:eom2}, but
could also include gravitational redshift or deflection in the
supernova context~\cite{Yang:2017asl}.
For neutrinos, the background medium causes a
modification of the refractive index and thus of ${\sf H}$, whereas
coherent forces by medium gradients are usually neglected and not even
mentioned.  Of course, a neutrino produced off-center in the Sun or in
a supernova core suffers refractive deflection unless it moves
radially, but this effect is extremely small compared with
gravitational deflection.  Likewise, neutrino diffraction e.g.\ by the
roughness of the Earth surface plays no practical role. For neutrinos,
only gravitational effects seem to be of any relevance for the
momentum drift term.

The doubts voiced in reference \cite{Hansen:2016klk} actually concern
the drift term in coordinate space, where $\bv$ is the velocity
corresponding to momentum $\bp$, i.e., $\bv=\bp/E_\bp$ where
$E_\bp^2=\bp^2+m^2$ for unmixed particles.  In the absence of
collisions or coherent forces, $\partial_t
f_{t,\bx,\bp}+\bv\cdot{\partial}_\bx f_{t,\bx,\bp}=0$ simply
represents conservation of particles with momentum $\bp$ (flux
conservation). In the context of mixed neutrinos, often the
ultra-relativistic limit was invoked to write flux conservation in the
form $\partial_t \varrho_{t,\bx,\bp}+\hat\bp\cdot{\partial}_\bx
\varrho_{t,\bx,\bp}=0$, where $\hat\bp$ is a unit vector in the
direction of $\bp$ \cite{Rudzsky:1990,Cardall:2007zw,Sirera:1998ia}.
In other words, propagation with the speed of light was assumed for
all $\varrho_{\bp}$ components. Non-vanishing masses were only
included in the oscillation matrix $\sH$.

While ignoring neutrino masses everywhere except in $\sH$ is probably
a good approximation in practice, we recall that the kinetic equation
derived by two of us a long time ago actually stated the Liouville
equation explicitly in the form~\cite{Sigl:1992fn}
\begin{equation}\label{eq:eom4}
\partial_t\varrho
+\textstyle{\frac{1}{2}}
\left\lbrace{\partial}_\bx\varrho,{\partial}_\bp \sH\right\rbrace
-\textstyle{\frac{1}{2}}
\left\lbrace{\partial}_\bp\varrho,{\partial}_\bx \sH\right\rbrace
= -i\[\sH,\varrho\]+{\cal C}\[\varrho\]\,,
\end{equation}
where both $\varrho$ and $\sH$ depend on $t$, $\bx$ and $\bp$ and
$\{{\cdot}\,,{\cdot}\}$ is an anti-commutator.  Notice that the
Hermitean matrix $\sH_{t,\bx,\bp}$ in diagonal form gives us the
energies of quanta with momentum $\bp$ in the medium which has
properties that can depend on $t$ and $\bx$. If the matter effect
contained in $\sH$ is the usual electroweak potential and as such does
not depend on $\bp$, the only part of $\sH$ that depends on $\bp$ is
the neutrino kinetic energy.  In this case
${\bm\sV}_\bp={\partial}_\bp \sH_{t,\bx,\bp}$ is a matrix of
velocities which is diagonal in the mass basis and has
$\bv_i=\bp/(\bp^2+m_i^2)^{1/2}$ appropriate for momentum $\bp$ on the
diagonal.  So we may interpret $\half\{\varrho,{\bm\sV}\}$ as a matrix
of neutrino fluxes and, in the absence of coherent forces or
collisions, equation~\eqref{eq:eom4} is simply flux conservation in
the form $\partial_t\varrho_{t,\bx,\bp}
+\half{\partial}_\bx\cdot\{{\bm\sV}_\bp,\varrho_{t,\bx,\bp}\}=0$.
For unmixed neutrinos this is flux conservation in the usual sense for
every separate mass eigenstate.

While the full Liouville term was stated in
reference~\cite{Sigl:1992fn}, we acknowledge that the derivation was
limited to the statement ``as one can easily show ...''---the main
focus at that time was the collision term and its impact on flavor
evolution.  In a more formal covariant derivation, Yamada~\cite{Yamada:2000za} found the
same structure (cf.\ his equation~35) but then turned quickly to the
ultra-relativistic limit (cf.\ text after equation~53). The same
structure is also borne out by references~\cite{Vlasenko:2013fja}
(cf.\ for example their equations 163--166), but was not spelled out
or discussed from a phenomenological perspective.
Reference~\cite{Sirera:1998ia} probably implies the same results,
although in their more phenomenological section they immediately turn
to the ultra-relativistic approximation (cf.\ their equation~73).  On
the other hand, the phenomenological derivation of the Liouville term
in reference~\cite{Cardall:2007zw} does not address the flavor
structure and applies only in the ultra-relativistic
limit.\footnote{The oscillation equations found here by two methods
  are of the form $p^\mu\frac{\partial}{\partial
    x^\mu}\,\varrho_{t,\bx,\bp} =-\frac{i}{2}\,[\sf
    M^2,\varrho_{t,\bx,\bp}]$, where the meaning of $p^0$ remains at
  first unspecified. Later the ultra-relativistic limit is taken so
  that $p^\mu=(|\bp|,\bp)$.}

This situation motivates us to return to this topic and provide in
section~\ref{sec:Liouville} a derivation of the Liouville term of
equation~\eqref{eq:eom4} with as little theoretical overhead as
possible. In section~\ref{sec:example} we work out an explicit
two-flavor example when keeping the flavor structure of the Liouville
term. We conclude with a summary and discussion in
section~\ref{sec:conclusion}.

\section{Spatial transport of mixed neutrinos}
\label{sec:Liouville}

\subsection{Derivation of the Liouville term}

\subsubsection{Two-point correlators}

On the level of a kinetic treatment our goal is to understand the
space-time evolution of the neutrino mean field in the form
of the $\varrho_{t,\bx,\bp}$ matrices which are expectation values
of field bilinears. The information contained in $\varrho_{t,\bx,\bp}$
is sufficient for all common questions arising in flavor
oscillation physics. First, they provide the refractive effect of
neutrinos on other neutrinos. Second, $\varrho_{t,\bx,\bp}$
allows us to calculate local interaction rates either in
astrophysical environments or in laboratory detectors.
Field bilinears are the lowest-order field correlators in the
Bogoliubov-Born-Green-Kirkwood-Yvon (BBGKY)
hierarchy~\cite{Volpe:2013jgr} and as such the lowest-order terms
of a systematic
perturbative expansion. Of course, there can be physical
circumstances where higher-order correlators are important, although
in astrophysical or laboratory neutrino physics
no such cases seem to have emerged.

Some authors prefer to study flavor oscillations in terms of
individual neutrinos propagating as wavepackets
\cite{Hansen:2016klk,Akhmedov:2009rb,Akhmedov:2017mcc,Giunti:2007ry}, a philosophy that
explicitly goes beyond the mean-field level. However,
as long as we only ask mean-field questions
(neutrino-neutrino refraction or local interaction rates)
this treatment does not produce new results. Moreover, it
requires unavailable information about the production
of the assumed wavepackets for which
only back-of-the-envelope estimates exist. Therefore, while
a kinetic treatment in terms of $\varrho_{t,\bx,\bp}$
matrices is of course not a complete description of the fluctuating neutrino gas
in a supernova or of the neutrino flux from a laboratory source,
this treatment is complete on the level of those questions
that are addressed in present-day neutrino physics.

The starting point for deriving the kinetic equation is the
Dirac quantum field\footnote{We denote second-quantized operators with
  a caret.}  $\hat\psi_i(t,\bx)$ that destroys a neutrino or
creates an antineutrino of flavor $i$ at time $t$ and location $\bx$.
This field and its conjugate are expanded in spatial
Fourier modes $\bp$ in terms of Dirac spinors and neutrino and
antineutrino destruction and creation operators $\hat a_i(h,\bp,t)$,
$\hat a_i^\dagger(h,\bp,t)$, $\hat b_i(h,\bp,t)$ and
\smash{$\hat b_i^\dagger(h,\bp,t)$}, where $h$ is the helicity. For
propagation in an isotropic medium
we only consider negative-helicity neutrinos and positive-helicity
antineutrinos. Moreover, we here focus on the advection part of the
kinetic equation\footnote{This terminology \cite{Yamada:2000za}, that
  is common in the supernova context, refers to the Liouville and
  flavor oscillation terms. In other words, it refers to phenomena
  caused by collisionless propagation.}  so that neutrinos and
antineutrinos are not coupled, for example, by pair processes.  We
thus consider a simplified model that includes only neutrino
destruction and creation operators
$\hat a_i(\bp,t)$ and $\hat a_i^\dagger(\bp,t)$
fulfilling the equal-time anti-commutation relation
\smash{$\{\hat a_i(\bp,t),\hat a_j^\dagger(\bp',t)\}=
(2\pi)^3\delta^3(\bp{-}\bp')\delta_{ij}$}. The advection part of the
Liouville equation for antineutrinos is the same except for a
well-known sign change in the refractive term.
After dismissing the entire
Dirac structure we could also consider bosons and use commutation
relations for the destruction and creation operators instead.

In the mean-field approximation, the system is described by
expectation values of field bilinears of the type
$\hat\psi_i^\dagger(t,\bx)\hat\psi_j(t,\bx)$. On the level of the
Fourier components we thus require expectation values of expressions
such as
\begin{equation}\label{eq:D-definition}
\hat \sD_{ij}(\bp,\bp',t)=\hat a_j^\dagger(\bp',t)\, \hat a_i(\bp,t)\,.
\end{equation}
As discussed earlier~\cite{Sigl:1992fn}, we dismiss fast-varying
bilinears of the type $\hat a_j^\dagger(\bp',t)\hat
a_i^\dagger(\bp,t)$ and also mixed bilinears between neutrinos and
antineutrinos, although in a non-isotropic medium,
neutrino-antineutrino pair correlations can be relevant after all
\cite{Serreau:2014cfa,Kartavtsev:2015eva}. If the medium is
homogeneous, the expectation value of every
observable constructed from the fields $\hat\psi_i$ and
\smash{$\hat\psi_j^\dagger$} is independent of
location, implying that the expectation value of
$\hat \sD_{ij}(\bp,\bp',t)$ contributes only at equal momenta.
Therefore, the mean field of a homogeneous neutrino
gas is completely characterized by dimensionless $N{\times}N$
``matrices of densities'' $\varrho_{\bp,t}$ given by
\smash{$\big\langle \hat a_j^\dagger(\bp',t)\, \hat a_i(\bp,t)\big\rangle=
  (2\pi)^3\delta^3(\bp-\bp')\,\(\varrho_{\bp,t}\)_{ij}$}.
The diagonal entries of $\varrho_{\bp,t}$ are the usual occupation
numbers of different flavors.

\subsubsection{Wigner transformation}

If the neutrino gas is inhomogeneous it is described by matrices
$\varrho_{t,\bx,\bp}$ that also depend on location.
Such a quasi-probability distribution in phase space makes
sense when inhomogeneities are weak, i.e.,
spatial variations are on scales much larger than a typical
neutrino wavelength. To arrive at this construction we make use
of the Wigner transformation
\cite{Wigner:1932eb,Hillery:1983ms,Lee:1995} and its inverse.
For a function $F(\bk,\bk')$ of two momentum variables it is
\begin{subequations}\label{eq:Wigner}
\begin{eqnarray}
  \tilde F(\bx,\bp)&=&\int\frac{d^3\bD}{(2\pi)^3}\,e^{i\bD\cdot\bx}\,
  F\big(\bp-{\textstyle\frac{\bD}{2}},\bp+{\textstyle\frac{\bD}{2}}\big)\,,
  \\
  F(\bk,\bk')&=&\int d^3\bx\,e^{-i(\bk'-\bk)\cdot\bx}\,
  \tilde F\big(\bx,\textstyle{\frac{\bk+\bk'}{2}}\big)\,.
\end{eqnarray}
\end{subequations}
With $\Delta=\bk'-\bk$ and $\bp=\half(\bk'+\bk)$
the inverse transformation can also be written as
\begin{equation}\label{eq:Wigner-inverse}
  F\!\(\bp-{\textstyle\frac{\bD}{2}},\bp+{\textstyle\frac{\bD}{2}}\)=
  \int d^3\bx\,e^{-i\bD\cdot\bx}\,\tilde F(\bx,\bp)\,.
\end{equation}
Such transformations are motivated in situations when $F(\bk,\bk')$ has
most of its power near $\bk=\bk'$ so that it makes sense to
use an average momentum. Moreover, in this case
$\tilde F(\bx,\bp)$ varies slowly as a function of $\bx$. However,
the Wigner transformation is a general mathematical operation that is not
limited to these assumptions.

Next we apply the Wigner transformation to the second-quantized
correlator $\hat\sD(\bk,\bk',t)$ defined in equation~\eqref{eq:D-definition},
\begin{equation}\label{Wigner2}
  \hat\varrho_{ij}(t,\bx,\bp)=
  \int\frac{d^3\bD}{(2\pi)^3}\,e^{i\bD\cdot\bx}\,
  \hat a_j^\dagger\!\(\bp-{\textstyle\frac{\bD}{2}},t\)\,
  \hat a_i\!\(\bp+{\textstyle\frac{\bD}{2}},t\)\,.
\end{equation}
Both $\hat \sD(\bk,\bk',t)$ and $\hat\varrho(t,\bx,\bp)$ carry the same
information. The mean field of the neutrino gas is finally
characterized by
$\varrho_{t,\bx,\bp}=\langle \hat\varrho_{t,\bx,\bp}\rangle$,
playing the role of space-varying occupation-number matrices.

\subsubsection{Equations of motion}

To derive equations of motion we begin with
Heisenberg's equation for an operator $\hat A$ in the form
$i\partial_t \hat A=[\hat A,\hat H]$, where $\hat H$ is the Hamiltonian.
We write it in the form
\begin{equation}\label{eq:hamiltonian}
  \hat{H}=\int\frac{d^3\bp}{(2\pi)^3}\,\frac{d^3\bp'}{(2\pi)^3}\,
  \hat a_i^\dagger(\bp)\,\sH_{ij}(\bp,\bp')\,\hat a_j(\bp')\,,
\end{equation}
where $\sH_{ij}$ is a matrix of numbers. Here and henceforth we no
longer show the dependence on time explicitly. A summation over
repeated flavor indices $i,j=1,\ldots,N$ is implied. This bilinear
form does not include neutrino-neutrino refraction that we leave out
for simplicity.

If the background medium is homogeneous, the matrix of energies
depends only on one momentum and is of the form
$\sH(\bp,\bp')=(2\pi)^3\delta^3(\bp{-}\bp')\,\sH^0(\bp)$. In particular,
in the mass basis this includes the energies
$(\bp^2+m_i^2)^{1/2}$ on the diagonal. On this level, the Hamiltonian $\hat H$
simply represents a collection of quantum harmonic oscillators. In addition,
there are refractive energy shifts that are diagonal in the interaction
basis. In the propagation basis, $\hat H$ still represents a collection of
harmonic oscillators. In general, however, the medium is not homogeneous
so that $\hat H(\bp,\bp')$
can deflect neutrinos, i.e., destroy one with momentum
$\bp'$ and create one with $\bp$.
This process is not a microscopic collision,
but rather a refractive deflection by weak inhomogeneities.

The evolution of the annihilation operator following from Heisenberg's equation is
given by
$i\partial_t\hat a_i(\bp)=[\hat a_i(\bp),\hat H(\bp,\bp')]=
\int d^3\bp'/(2\pi)^{3}\,\,\sH_{ik}(\bp',\bp)\,\hat a_k(\bp')$. With this
result we can evaluate Heisenberg's equation for $\hat\varrho(\bx,\bp)$ and find
\begin{eqnarray}
i\partial_t \hat\varrho_{ij}(\bx,\bp)&=&
\int\frac{d^3\bD}{(2\pi)^3}\,\frac{d^3\bp'}{(2\pi)^3}\,e^{i\bD\cdot\bx}\,
\Bigl[\sH_{ik}\!\(\bp+\textstyle{\frac{\bD}{2}},\bp'\)\,
\hat a_j^\dagger\!\(\bp-\textstyle{\frac{\bD}{2}}\)\, \hat a_k\!\(\bp'\)
\nonumber\\
&&\kern9em{}-
\hat a_k^\dagger\!\(\bp'\)\,\hat a_i\!\(\bp+\textstyle{\frac{\bD}{2}}\)\,
\sH_{kj}\!\(\bp-\textstyle{\frac{\bD}{2}},\bp'\)\Bigr]\,.
\end{eqnarray}
We next introduce the variables $\bD_1$ and $\bD_2$ that are defined by
$\bp'=\bp+\half(\bD_1-\bD_2)$ and $\bD=\bD_1+\bD_2$, leading to the more
symmetric expression
\begin{eqnarray}
i\partial_t \hat\varrho_{ij}(\bx,\bp)&=&
\int\frac{d^3\bD_1}{(2\pi)^3}\,\frac{d^3\bD_2}{(2\pi)^3}\,e^{i(\bD_1+\bD_2)\cdot\bx}\,
\Bigl[\sH_{ik}\!\(\bp_1{+}\textstyle{\frac{\bD_2}{2}},\bp_1{-}\textstyle{\frac{\bD_2}{2}}\)\,
\hat a_j^\dagger\!\(\bp_2{-}\textstyle{\frac{\bD_1}{2}}\)\,
\hat a_k\!\(\bp_2{+}\textstyle{\frac{\bD_1}{2}}\)
\nonumber\\
&&\kern11.2em{}
-\hat a_k^\dagger\!\(\bp_1{-}\textstyle{\frac{\bD_2}{2}}\)\,
\hat a_i\!\(\bp_1{+}\textstyle{\frac{\bD_2}{2}}\)\,
\sH_{kj}\!\(\bp_2{-}\textstyle{\frac{\bD_1}{2}},\bp_2{+}\textstyle{\frac{\bD_1}{2}}\)
\Bigr]\,,
\nonumber\\
\end{eqnarray}
where we have used the notation $\bp_1=\bp+\half\bD_1$ and $\bp_2=\bp-\half\bD_2$.
Notice that the integrals over $d^3\bD_{1,2}$ cannot be evaluated
to produced Wigner
transforms because $\bD_{1,2}$ is also hidden in $\bp_{1,2}$. However,
under the integral we can substitute for each factor the
inverse Wigner transformation in the form of equation~\eqref{eq:Wigner-inverse}
and find
\begin{eqnarray}\label{eq:commutator-1}
i\partial_t \hat\varrho_{\bx,\bp}&=&
\int\frac{d^3\bD_1}{(2\pi)^3}\,\frac{d^3\bD_2}{(2\pi)^3}\,d^3\bx_1\,d^3\bx_2\,
e^{-i\bD_1\cdot(\bx_1-\bx)-i\bD_2\cdot(\bx_2-\bx)}
\,\Bigl[\sH_{\bx_2,\bp_1}\,\hat\varrho_{\bx_1,\bp_2}
-\hat\varrho_{\bx_2,\bp_1}\,\sH_{\bx_1,\bp_2}\Bigr]\,,
\nonumber\\
\end{eqnarray}
where we have used $\bx_{1,2}$ as the conjugate variables to $\bD_{1,2}$.

To obtain the argument $\bp$ instead of $\bp_{1,2}$ we use the shift
operator in the form $F(\bk+\bq)=e^{\bq\cdot\partial_\bk}\,F(\bk)$.
This construction implies e.g.\
\smash{$\sH(\bx_2,\bp_1)=e^{\frac{1}{2}\bD_1\cdot\partial_\bp}\sH(\bx_2,\bp)$}
and overall we find
\begin{eqnarray}\label{eq:commutator-2}
\kern-2.5em i\partial_t \hat\varrho_{\bx,\bp}&=&
\int\frac{d^3\bD_1}{(2\pi)^3}\,\frac{d^3\bD_2}{(2\pi)^3}\,d^3\bx_1\,d^3\bx_2\,
\Bigl[\sH_{\bx_2,\bp}
\,e^{-i\bD_1\cdot(\bx_1-\bx+\frac{i}{2}\leftpartial_\bp)
-i\bD_2\cdot(\bx_2-\bx-\frac{i}{2}\rightpartial_\bp)}
\,\hat\varrho_{\bx_1,\bp}
\nonumber\\
&&\kern10.6em{}-\hat\varrho_{\bx_2,\bp}
\,e^{-i\bD_1\cdot(\bx_1-\bx+\frac{i}{2}\leftpartial_\bp)
-i\bD_2\cdot(\bx_2-\bx-\frac{i}{2}\rightpartial_\bp)}
\,\sH_{\bx_1,\bp}\Bigr]\,,
\end{eqnarray}
where $\leftpartial_\bp$ means that the differential operator is to
be applied to the expression left of it. Using the representation of the delta function
$\delta^{(3)}(\bx)=\int d^3\bD e^{i\bD\cdot\bx}/(2\pi)^3$ it is now straightforward to
evaluate the integrals\footnote{The $\delta$-function of a momentum derivative arising in equation \eqref{eq:commutator-2} can be avoided. This is demonstrated by a simplified version of equation \eqref{eq:commutator-1},
\begin{flalign}
	\int d x_1 \frac{d \Delta}{2 \pi} e^{- i \Delta \( x_1 - x \)} f \( p+\Delta \) g \( x_1 \)
		=& \int d x_1 \frac{d \Delta}{2 \pi} e^{- i \Delta \( x_1 - x \)} \sum_{n=0}^{\infty} \frac{1}{n!} \( \Delta \partial_p \)^n f (p) g\(x_1\)
		\nonumber\\
		=& \sum_n \frac{1}{n!} \int d x_1 \frac{d \Delta}{2 \pi} e^{- i \Delta \( x_1 - x \)} \( i \overleftarrow{\partial}_{x_1 - x} \overrightarrow{\partial}_p \)^n f(p) g\(x_1\)
		\nonumber\\
		=& \sum_n \frac{i^n}{n!} f^{(n)} (p) \int d x_1
                \delta^{(n)} \( x_1-x \) g\(x_1\).
                \nonumber
\end{flalign}
After an integration by parts it is obvious that a $\delta$-function of a derivative leads to the same result as a sum of derivatives of a $\delta$-function in this specific context.} and, for example, the first term in square brackets
becomes $\sH\bigl(\bx+\frac{i}{2}\rightpartial_\bp,\bp\bigr)\,
\hat\varrho\bigl(\bx-\frac{i}{2}\leftpartial_\bp,\bp\bigr)$. The differential operator in the argument of one matrix is to be applied to the other matrix.
A more elegant way to express this structure is found by using once more the
shift operator to lift the deviation from $\bx$ in the arguments to an
exponential,
\begin{equation}\label{eq:commutator-3}
i\partial_t \hat\varrho_{\bx,\bp}=
\sH_{\bx,\bp}
\,e^{\frac{i}{2}(\leftpartial_\bx\cdot\rightpartial_\bp
-\leftpartial_\bp\cdot\rightpartial_\bx)}
\,\hat\varrho_{\bx,\bp}
-\hat\varrho_{\bx,\bp}
\,e^{\frac{i}{2}(\leftpartial_\bx\cdot\rightpartial_\bp
-\leftpartial_\bp\cdot\rightpartial_\bx)}
\,\sH_{\bx,\bp}\,.
\end{equation}
An equation equivalent to this result was first derived by
Moyal~\cite{Moyal:1949} with two minor differences. We here use
matrices in flavor space as opposed to scalar functions and our matrix
$\hat\varrho$ is a second-quantized operator as opposed to a purely
quantum-mechanical setting.  We also note that if we were to keep
Planck's constant $\hbar$, it would multiply the lhs of
equation~(\ref{eq:commutator-3}) as well as the exponents on the right
hand side (rhs).

\subsubsection{Mean field}

Up to this point we have not made any approximations in that
equation~\eqref{eq:commutator-3} follows from Heisenberg's equation
for $\hat\varrho$ under the Hamiltonian $\hat H$ defined in
equation~\eqref{eq:hamiltonian}. Next we take the expectation value of
$\hat\varrho$ so that we substitute
$\hat\varrho_{\bx,\bp}\to\varrho_{\bx,\bp}$.  If we assume that
$\sH_{\bx,\bp}$ and $\varrho_{\bx,\bp}$ vary only slowly as a function
of their arguments we may expand equation~\eqref{eq:commutator-3} to
lowest order, providing the advection part of
equation~\eqref{eq:eom4}, i.e.\ the Liouville term, now for the
nontrivial matrix structure in flavor space, and the refractive
term. It is interesting to note that if one would keep $\hbar$
explicitly, it would cancel out in the Liouville term, consistent with
its classical nature.

Actually the advection term can also be found more directly from
equation~\eqref{eq:commutator-1}. We can take the expectation value
$\hat\varrho_{\bx,\bp}\to\varrho_{\bx,\bp}$ already in that equation,
assume that  $\sH_{\bx,\bp}$ and $\varrho_{\bx,\bp}$
vary only slowly as a function of their arguments, expand them
and perform the integrals.

The diagonal elements of the mean field $\varrho_{\bx,\bp}$ are
space-dependent occupation numbers. Because of the quantum-mechanical
uncertainty between $\bx$ and $\bp$ this concept makes sense only if
the spatial variations are slow compared with a typical neutrino
wavelength. It is known that Wigner functions $f_{\bx,\bp}=\langle
\hat f_{\bx,\bp}\rangle$, here for unmixed particles, are not
guaranteed to be non-negative so that the interpretation as a
probability distribution in phase space is not obvious. On the other
hand, if $f_{\bx,\bp}$ is coarse-grained over phase-space regions
corresponding to the uncertainty relation it is non-negative
\cite{Hillery:1983ms,Lee:1995}.  One way to implement this idea is the
Husimi transformation \cite{Husimi:1940,Cartwright:1976} that uses a
Gaussian smearing of the Wigner distribution. We show in
appendix~\ref{app:Husimi} that the Husimi distribution leads to the
same Liouville term up to corrections of the order of the implemented
phase-space blurring.

However, the question if a quantity like $\langle \hat f_{\bx,\bp}\rangle$
is non-negative appears to be moot for the kinetic
equation~\eqref{eq:eom4}. This is a closed set of differential
equations where the $\varrho$ matrices are not found from taking
expectation values of underlying quantum operators, but the
local occupations of modes
are filled or depleted by source, sink and collision
terms as well as free streaming. We are not aware that this kinetic
equation could produce pathological solutions such as
negative occupation numbers. In other words, we have
formulated the kinetic equation in terms of
Wigner functions, but the solutions of this equation
are independent of the true underlying quantum system. There is no
guarantee that the kinetic equation produces the same results that
would be found by solving the full quantum system, but the kinetic
equation itself appears to be well behaved. Of course, we expect
that the solution of the kinetic equation agrees with the full
system unless we consider scales where the
uncertainty relation is important, but on such scales one
would not use the kinetic equation anyway, at least not in
the context of neutrino transport and flavor oscillations.

\subsection{Particle transport or wave equation?}

The mean-field description in terms of $\varrho_{t,\bx,\bp}$
naturally obeys a partial differential equation with independent
derivatives in all arguments. Some of the doubts about
this structure \cite{Hansen:2016klk} are apparently related to
picturing the evolution of $\varrho_{t,\bx,\bp}$ as describing a
single neutrino on a trajectory so that the time and space variables
are said to be related by a classical trajectory of the type
$\bx=\bv\,t$. Related to this doubt is apparently the question of the
connection between a partial differential operator on the lhs of the
kinetic equation with an ordinary differential equation of the
form~\eqref{eq:eom0} for flavor oscillations. Somewhat in reverse,
Cardall~\cite{Cardall:2007zw} started from the picture of flavor
oscillations along a trajectory of the form~\eqref{eq:eom0} and argued
in two different ways on how to arrive at a Liouville equation
$(\partial_t+\hat\bp\cdot{\partial}_\bx)\varrho_{t,\bx,\bp}$, where
$t$ and $\bx$ are independent variables.

One simple heuristic way to understand the appearance of the Liouville
operator is to ignore the collision part of the kinetic equation as in
references~\cite{Hansen:2016klk,Cardall:2007zw} and focus on flavor
oscillations alone. However, in this case we do not need matrices of
densities and can formulate the problem on the level of wave
amplitudes as in equation~\eqref{eq:eom0}. Flavor oscillations and
similar phenomena arise from the interference of wave components with
different dispersion relations and as such are wave phenomena. If we
ignore issues of helicity or particle-antiparticle oscillations for
neutrinos we may ignore the Dirac structure and in vacuum each field
component obeys the Klein-Gordon equation
$(\partial_t^2-{\partial}_\bx^2)\psi_i=-m_i^2\psi_i$ for
$i=1,\ldots,N$. For one space dimension this is
\begin{equation}\label{eq:eom5}
  (\partial_t-\partial_x)(\partial_t+\partial_x)\psi=-\sM^2\psi\,,
\end{equation}
where $\sM^2={\rm diag}\,(m_1^2,\ldots,m_N^2)$ in the mass basis.
Next we consider a plane-wave solution of the form
$e^{-i(Et-\bp\cdot\bx)}$, which however involves a different $\bp_i$
for every $m_i$ for a common $E$. However, in the ultra-relativistic
limit $E\approx|\bp|$ and we can linearize the Klein-Gordon equation
if we observe that for such plane waves $(\partial_t-\partial_x)\to
-i(E+p_x) \approx -2 i E$. In the second factor this approximation
would not be possible because what appears is the difference between
$E$ and $p_x$ that would vanish in the same approximation. Essentially
by separating the scales between the flavor oscillation length and the
neutrino wavelength we arrive at
\begin{equation}\label{eq:eom6a}
  (\partial_t+\partial_x)\psi=-i\frac{\sM^2}{2E}\,\psi\,.
\end{equation}
Of course, we can also consider the complex conjugate equation and
combine them in the usual way as an equation for the density matrix
$\rho_{ij}=\psi^*_j\psi_i$ in the form
\begin{equation}\label{eq:eom6b}
  (\partial_t+\hat\bp\cdot{\partial}_\bx)\rho=-i[\sH,\rho]\,,
\qquad\hbox{where}\qquad \sH=\frac{\sM^2}{2E}
\end{equation}
and we have restored a general direction $\hat\bp$ of propagation.

The practical meaning of equation~\eqref{eq:eom6b} depends on initial
and/or boundary conditions. If we consider a homogeneous situation
without spatial gradients, we are back to a simple Schr\"odinger
equation of the form $\partial_t\rho=-i[\sH,\rho]$. If we consider a
stationary source, nothing depends on time and we need to solve
$\partial_x\rho=-i[\sH,\rho]$ along the beam. Either way, such an
equation only applies to monochromatic waves because $\sH$ is only
defined for a specific $E\approx|\bp|$. Of course one can also
consider wavepackets, but then an equation of the
form~\eqref{eq:eom6a} needs to be solved for every Fourier
component. Therefore, a wavepacket has $\rho$ matrices involving wave
amplitudes with different $E$ values, not only those with equal
$E$. However, in all practical oscillation experiments we take the
average over many measured neutrinos (each of which may have been
emitted as a wavepacket) and in this average the phase relations
between different Fourier components of individual wavepackets are
lost.  Therefore, such an ensemble average requires only the
occupation numbers $\varrho_\bp$ of the beam, not the phase relations
between Fourier components encoded in
wavepackets~\cite{Stodolsky:1998tc}.

So what we make of equations~\eqref{eq:eom6a} and~\eqref{eq:eom6b}
depends on the specific physical circumstances and on the questions we
wish to address. Either way, these equations can be seen as wave
equations in the approximation of ultra-relativistic neutrinos.  They
do not require or motivate an interpretation in terms of point-like
particles moving along classical trajectories. In particular, in the
stationary-source example, there is no need to think of a neutrino
with momentum $\bp$ to exist at some precise location $\bx$ in
violation of Heisenberg's uncertainty relation. All we need is a
boundary condition at some location, not a localization of the wave
itself. The physical nature of the boundary condition is not part of
the wave equation or of the Liouville operator.

In astrophysics, a kinetic equation of unmixed particles is used, for
example, as the basis for treating radiative transport by photons or
neutrinos. Particle fluxes are driven by gradients of temperature or
lepton number. The mean free path of the particles and the relevant
gradients are large compared with the radiation wavelength. Then the
wave nature of the radiation is irrelevant and one may think of the
kinetic equation as describing classical particles.  However, contrary
to the doubts voiced in reference~\cite{Hansen:2016klk}, the kinetic
equation is not an equation for point-like classical
particles. Rather, we only use it on scales where the distinction
between waves and particles is irrelevant.  The kinetic equation does
not assume that neutrinos are localized within phase space to better
than allowed by the uncertainty relation. The function
$\varrho_{\bx,\bp}$ does not give us the flavor content of a specific
neutrino that would be localized precisely at $(\bx,\bp)$ in phase
space.  The mean field $\varrho_{\bx,\bp}$ does not describe the
localization and flavor content of individual particles. Rather it
describes the expectation values of occupation number operators in
flavor space which themselves depend on location and momentum.

Once we include flavor oscillations the wave nature of the underlying
radiation becomes apparent and we can interpret the advection part of
the kinetic equation as a first-order wave equation as argued
earlier. Indeed, the derivation shown earlier is entirely quantum
physical. It is only the final step of taking expectation values and
the subsequent Taylor expansion where small-scale information is lost.

The interpretation of the advection part of the kinetic
equation~\eqref{eq:eom4} as a wave equation is crucial in the context
of self-induced flavor conversion by neutrino-neutrino refraction
\cite{Pantaleone:1992eq,Duan:2010bg}. In this situation the neutrino
mean field acts back on itself through the oscillation term, i.e.,
$\sH$ depends on the collection of $\varrho$ matrices.  Without this
effect, flavor conversion is a purely kinematical phenomenon that
arises from the interference of independently propagating waves that
do not know about each other. Neutrino-neutrino refraction causes
these ``flavor waves'' to become dynamical and we obtain a first-order
wave equation with propagating and/or run-away solutions
\cite{Izaguirre:2016gsx,Capozzi:2017gqd}.


\section{Flavor-dependent Liouville term: Phenomenological consequences}
\label{sec:example}

\subsection{Matrices of velocities}

To develop some phenomenological understanding of the flavor-dependent
Liouville term in the kinetic equation~\eqref{eq:eom4} we ignore the
momentum drift term caused by external forces. It appears to be
dominated by gravitational effects in all practical cases. We also
ignore the collision term and thus only worry about the advection part
without external forces. We recall that in an isotropic dispersive
medium, a wave with frequency $\omega$ and wavevector $k=|\bk|$ has
group velocity $\partial\omega/\partial k$ and phase velocity
$\omega/k$.  In this sense ${\bm\sV}_\bp=\partial_\bp\sH$ is a matrix
of group velocities whereas $\sH$, after dividing by $|\bp|$, is a
matrix of phase velocities.  Flavor oscillations are an interference
effect between waves with different dispersion relations and so it
comes as no surprise that the oscillation term involves phase
velocities.

Particles with mass have energy $E=(\bp^2+m^2)^{1/2}$ so that
the group velocity is $\partial E/\partial p=p/E\approx 1-m^2/2 p^2$
with $p=|\bp|$. Of course, this is the usual particle velocity and thus
subluminal. The phase velocity $E/p\approx 1+m^2/2 p^2$, on the other
hand, is superluminal. Therefore, in the ultra-relativistic limit,
the two velocities deviate from the speed of light by the same amount
in opposite directions.

For mixed neutrinos, the group velocities appear on the lhs of the kinetic equation, the phase velocities on the rhs. In practice we always consider
ultra-relativistic neutrinos, so all group and phase velocities are
very close to the speed of light. However, the phase velocities appear
in a commutator, i.e., it is the difference between phase velocities
that causes flavor oscillations, whereas the group velocities appear
in an anti-commutator. Therefore, to lowest order in the small
deviation from the speed of light we may use
${\bm\sV}_\bp\approx\hat\bp$, an approximation that was always used in
the literature.\footnote{Instead of $\hat\bp$ one often
  used $\bv$ to denote a unit vector in the direction of
  $\bp$. However, in reference~\cite{Hansen:2016klk} and in our
  further discussion, $\bv$ stands for the average velocity of two
  neutrino mass eigenstates.}

In reference~\cite{Hansen:2016klk} the impact of having different
group velocities was studied in the context of a Liouville equation
for wavepackets and a new damping term was found. It was attributed to
the effect of wavepacket separation which leads to the loss of flavor
coherence. Of course, this effect is not new and has been studied many
times in the context of wavepacket discussions of flavor
oscillations. This loss of flavor coherence must be interpreted in the
sense of kinematical decoherence and as such is included in the
advection part of the kinetic equation~\eqref{eq:eom4}, whereas
dynamical decoherence from the entanglement with the environment
is caused by the collision term as mentioned earlier.  After some
distance of propagation, the off-diagonal elements of
$\varrho_{\bx,\bp}$ vary fast as a function of $|\bp|$ so that a
detector with insufficient energy resolution can no longer see the
oscillatory pattern. However, this apparent loss of flavor coherence
is related to the detector properties and should not be part of the
neutrino equation of motion.

\subsection{Stationary source}

While the deviation of the neutrino group velocities
from the speed of light is a
higher-order effect, it is still interesting to consider
a simple example for the impact
of the full ${\bm\sV}_\bp$ matrix on flavor oscillations. To this end
we consider a stationary situation (nothing depends on
time), i.e., a stationary neutrino source and we ask
for the flavor content as a function of distance from the source.
So we consider the equation
\begin{equation}\label{eq:kinetic-eq-1}
\textstyle{\frac{1}{2}}
\left\lbrace{{\bm\sV}_{\bp},\bm\nabla}_\bx\varrho_{\bp,\bx}\right\rbrace
= -i\[\sH_{\bp},\varrho_{\bp,\bx}\]\,,
\qquad\hbox{where}\qquad
{\bm\sV}_{\bp}={\bm\nabla}_\bp\sH_\bp\,.
\end{equation}
The background medium is taken to be homogeneous, isotropic and stationary
and we ignore neutrino-neutrino refraction. Therefore, in the
mass basis the matrix of velocities is simply
${\bm\sV}_{\bp}={\rm diag}(\bv_1,\ldots,\bv_N)$
with $\bv_i=\bp/(\bp^2+m_i^2)^{1/2}$.

One first observation concerns the conservation of particles, often
stated as conservation of ${\rm tr}(\varrho_{\bp})$. Indeed
the rhs of equation~\eqref{eq:kinetic-eq-1}
is traceless due to its commutator structure.
If we consider a time-dependent situation with $\partial_t\varrho_{\bp,t}$
on the lhs we see that indeed ${\rm tr}(\varrho_{\bp,t})$ is conserved.
However, in our case it is the trace of the flux matrix
$\half\lbrace{\bm\sV}_\bp,\varrho_{\bp,\bx}\rbrace$
which is conserved,
in agreement with physical intuition. If a source produces neutrinos
at a given rate, the flux through a surface surrounding the source
is stationary. As neutrinos with different mass propagate at different
speeds, the local neutrino density outside of the source depends on the
velocity. Slower-moving neutrinos take a longer time to cover the distance
between the source and detector and so their density must be larger.

\subsection{Two flavors}

Henceforth it is understood that ${\bm\sV}$ and $\sH$ depend on $\bp$ and
$\varrho$ on $\bp$ and $\bx$, so we no longer show these variables
explicitly. Moreover, we turn to a two-flavor system and write the
velocity matrix in the mass basis in the form
\begin{equation}\label{eq:velocity}
  {\bm\sV}=\begin{pmatrix}\bv_1&0\\0&\bv_2\end{pmatrix}
  =\bv\,\sigma_0+\frac{\delta\bv}{2}\,\sigma_3\,,
\end{equation}
where $\sigma_j$ ($j=0,\ldots,3$) are Pauli matrices with $\sigma_0$
the $2{\times}2$ unit matrix. Moreover, we use
$\bv=(\bv_1+\bv_2)/2$ and $\delta\bv=\bv_1-\bv_2$. Thus we need to solve
the equation
\begin{equation}\label{eq:kinetic-eq-2}
\bv\cdot{\bm\nabla}\varrho+\frac{\delta\bv}{4}\cdot
\left\lbrace{\bm\nabla}\varrho,\sigma_3\right\rbrace
=-i\[\sH,\varrho\]\,.
\end{equation}
Without loss of generality we consider a
one-dimensional system evolving in the $x$-direction and we use the notation
$\varrho'=\partial_x\varrho$. Therefore, we need to solve
\begin{equation}\label{eq:kinetic-eq-3}
  \varrho'+\delta_v\,\left\lbrace\varrho',\frac{\sigma_3}{2}\right\rbrace=
  -i\[\frac{\sH}{v},\varrho\]\,,
\end{equation}
where $v=(v_1+v_2)/2$ and $\delta_v=(v_1-v_2)/(v_1+v_2)$. The appearance
of $\sH/v$ on the rhs is understood
because with $v=|\bp|/E$ we notice that
$\sH/v$ is something like a matrix of wave numbers, which is appropriate
for the phase evolution along a beam.

The meaning of this equation becomes more transparent if we write it in terms
of polarization vectors $\bB$ and $\bP$ defined by
\begin{equation}
\frac{\sH}{v}=\sum_{j=1}^3 B_j\frac{\sigma_j}{2}
\qquad\hbox{and}\qquad
\varrho=\sum_{j=0}^3 P_j\frac{\sigma_j}{2}\,.
\end{equation}
Notice that we write $\sH/v$ in traceless form because it always appears
in a commutator, whereas for $\varrho$ we include the trace
in terms of $P_0$.
Equation~(\ref{eq:kinetic-eq-3}) then takes the form
\begin{equation}\label{eq-kinetic-eq-4}
  P_0'+\delta_v\,P_3'=0
  \qquad\hbox{and}\qquad
  \bP'+\delta_v\,P_0'\,\be_3=\bB\times\bP\,,
\end{equation}
where $\be_3$ is the unit vector in the mass direction in flavor space.
Sticking the first
equation into the second provides
\begin{equation}\label{eq-kinetic-eq-5}
  P_0'=-\delta_v\,P_3'
  \qquad\hbox{and}\qquad
  \bP'-\delta_v^2\,P_3'\,\be_3=\bB\times\bP\,.
\end{equation}
Therefore, the three components of $\bP$ obey a closed set of
differential equations
\begin{equation}\label{eq-kinetic-eq-6}
  P_1'=B_2\,P_3-B_3\,P_2\,,
  \qquad
  P_2'=B_3\,P_1-B_1\,P_3\,,
  \qquad\hbox{and}\qquad
  P_3'=\frac{B_1\,P_2-B_2\,P_1}{1-\delta_v^2}\,.
\end{equation}
This equation is simplified with the notation
$\tilde P_3=P_3 \sqrt{1-\delta_v^2}$ and $\tilde B_{1,2}=B_{1,2}/\sqrt{1-\delta_v^2}$,
whereas for the other components the symbols with or without tilde are the same.
Then the equation of motion reads
\begin{equation}\label{eq-kinetic-eq-7}
  \tilde\bP'=\tilde\bB\times\tilde\bP\,.
\end{equation}
Therefore, the evolution is an ordinary precession of an abstract
polarization vector $\tilde\bP$ around an abstract magnetic field $\tilde\bB$.
The evolution is perfectly periodic---there is no damping.

It is the length of $\tilde\bP$ that is conserved, not the length
of $\bP$. On the other hand, if we prepare the system in an eigenstate
of $\sH$ (propagation eigenstate) then initially $\bP\propto\bB$ which
also implies that initially $\tilde\bP\propto\tilde\bB$. So
there are no oscillations along the beam in the same way
as there would be no oscillations in the corresponding
time-dependent problem.

\subsection{Interpretation}

In the simplest case of vacuum oscillations, the mass basis is
identical with the propagation basis and with the basis where the
velocity matrix is diagonal. In the two-flavor context of the previous
section this implies $B_{1,2}=0$ and thus $\tilde B_{1,2}=0$, i.e., the
polarization vectors precess around the mass direction in flavor
space. As a consequence, $P_3'=0$ and thus $P_0'=0$, i.e., the
projection of the polarization vector on the mass direction is
conserved and thus also $P_0$. Therefore, in this case
${\rm tr}(\varrho)$ remains conserved: along the beam both the
neutrino flux and the neutrino density remain the same.

This result makes physical sense. At the source we produce some
coherent combination of mass eigenstates which then propagate
independently. The probability for any mass eigenstate along the beam
remains constant, only their relative phases evolve, implying
oscillations of interaction eigenstates. Of course, the flux ratio
of the mass eigenstates is not the same as their density ratio, but
both remain constant along the beam.

In a dispersive medium, the group velocity of a wave
can be a complicated expression. In our case the background medium
produces a simple potential, i.e., the same shift of energy for
all $\bp$ so that for a given $\bp$ the group velocities of the particles are
the same with or without the potential. However, the matrix
of particle velocities is not diagonal in the same basis as
$\sH$ which is proportional to the matrix
of phase velocities. Therefore, along the beam we not only
have oscillations between interaction eigenstates but also
oscillations between mass eigenstates and thus between eigenstates of
velocity. Therefore, the overall particle density along the beam
cannot be the same if the overall particle flux is conserved.
Therefore, it makes physical sense that the evolution of
$\bP$, which describes the particle density, is not a simple
precession and that ${\rm tr}(\varrho)$, represented by
$P_0$, varies along the beam.

We finally notice, e.g.\ from equation~(\ref{eq-kinetic-eq-5}), that
the flavor dependence of neutrino velocities causes modifications of
the order of $\delta_v^2\ll1$, i.e., of the order of
$(m_1^2-m_2^2)^2/(2|\bp|)^4$. Therefore, these corrections are of
higher order compared with flavor oscillation effects as argued
earlier.

\section{Conclusion}
\label{sec:conclusion}

We have derived the advection part of the kinetic equation~\eqref{eq:eom4}
using only the most elementary ingredients of field theory. This
derivation fills a gap left in our earlier paper~\cite{Sigl:1992fn} and
is complementary to more recent derivations based on a more advanced
formalism. Our step-by-step derivation should be accessible to anyone
interested in flavor oscillations.

The matrix of neutrino velocities appearing in the Liouville
operator implies a conceptually interesting deviation from the
usual picture of flavor oscillations. As the neutrino
flux is conserved along the beam, periodic
modulations of the neutrino density can obtain.
However, the corrections
are of higher order in the small deviation of neutrino velocities from the
speed of light, the dominant term being the usual commutator expression
causing oscillations. Therefore, in situations
of practical interest we can use the standard approach of assuming
the speed of light for neutrinos everywhere except in the matrix $\sH$.

While the Liouville operator is identical to the transport part
of a kinetic equation for classical particles,
we have argued that the advection part of equation~\eqref{eq:eom4}
can be seen as a linearized first-order wave equation, in particular for
the flavor degree of freedom. Such an interpretation
becomes crucial in the presence of neutrino-neutrino refraction when
this equation can be seen as a dynamical equation for ``flavor waves''
with their own dispersion relation \cite{Izaguirre:2016gsx,Capozzi:2017gqd}.
With this interpretation,
equation~\eqref{eq:eom4} may ultimately lead to a better understanding
of neutrino flavor evolution in core-collapse supernovae or
neutron-star mergers.

\section*{Acknowledgments}

We acknowledge partial support by the Deutsche Forschungsgemeinschaft through
Grants No.\ EXC 153 (Excellence Cluster ``Universe''),
SFB 1258 (Collaborative Research Center ``Neutrinos, Dark Matter,
Messengers''), and SFB 676 (Collaborative Research Center ``Particles,
Strings, and the Early Universe'') as well as by the European Union through Grant
No.\ H2020-MSCA-ITN-2015/674896 (Innovative Training Network
``Elusives'').

\appendix

\section{Husimi distribution}
\label{app:Husimi}
Here we derive the Liouville equation from quantum mechanics in yet
another way, using the Husimi transformation~\cite{Husimi:1940}.  For
simplicity we restrict ourselves to one flavor. The smearing in
location and momentum space represented by the Husimi distribution can
be defined as
\begin{equation}\label{Husimi1}
  F(\bx,\bp)\equiv\frac{1}{(2\pi\eta\sigma)^3}\int d^3\bx^\prime d^3\bp^\prime f(\bx^\prime,\bp^\prime)
  \exp\left[-\frac{(\bx-\bx^\prime)^2}{2\eta^2}-\frac{(\bp-\bp^\prime)^2}{2\sigma^2}\right]\,,
\end{equation}
where $\eta$ and $\sigma$ are the length and momentum scales, respectively, over which the Wigner distribution
$f(\bx,\bp)$ is smeared and here and in the following we again suppress time dependencies.
To express the momentum integral in terms of a spatial integral it is useful to rewrite the second-quantized Wigner
distribution corresponding to equation~(\ref{Wigner2}) in terms of a spatial integral,
\begin{equation}\label{Wigner2a}
  \hat f(\bx,\bp)=\int\frac{d^3\bD_x}{(2\pi\hbar)^3}\,e^{i\bp\cdot\bD_x/\hbar}\,
  \hat\psi^\dagger\!\(\bx+{\textstyle\frac{\bD_x}{2}}\)\,
  \hat\psi\!\(\bx-{\textstyle\frac{\bD_x}{2}}\)\,,
\end{equation}
where we have kept $\hbar$ explicit and the spatial wave function
operator $\hat\psi(\bx)$ is related to the destruction operator $\hat
a(\bp)$ by
\begin{equation}\label{a_psi}
  \hat\psi(\bx)=\int\frac{d^3\bp}{(2\pi\hbar)^{3/2}}\,e^{-i\bp\cdot\bx/\hbar}\,\hat a(\bp)\,.
\end{equation}
Choosing $\sigma=\hbar/(2\eta)$, a straightforward calculation then
allows us to express the operator valued equivalent of
equation~(\ref{Husimi1}) as
\begin{equation}\label{Husimi2}
  \hat F(\bx,\bp)=\frac{1}{(2\pi\eta^2)^{3/2}}
  \int\frac{d^3\bx_1 d^3\bx_2}{(2\pi\hbar)^3}\hat\psi^\dagger(\bx_1)\hat\psi(\bx_2)
  \exp\!\[-\frac{(\bx{-}\bx_1)^2+(\bx{-}\bx_2)^2}{4\eta^2}
  +\frac{i\bp\cdot(\bx_1{-}\bx_2)}{\hbar}\].
\end{equation}
For the time evolution of the quantum field $\hat\psi$ we now make the ansatz
\begin{equation}\label{psi_t}
  i\hbar\partial_t\hat\psi=\bv\cdot\bp\hat\psi+V(\bx)\hat\psi\,,
\end{equation}
leading to
$\partial_t\hat\psi=-\bv\cdot\partial_\bx\hat\psi-iV(\bx)\hat\psi/\hbar$.
The obvious identity
$\bv\cdot(\partial_{\bx_1}+\partial_{\bx_2})\exp[\cdots]=-\bv\cdot\partial_\bx\exp[\cdots]$
and partial integration imply that the time
derivative $\partial_t\hat F(\bx,\bp)$ is given by the same integral
as in equation~(\ref{Husimi2}) with the additional operator
$-\bv\cdot\partial_\bx+i\left[V(\bx_1)-V(\bx_2)\right]/\hbar$ acting
on the exponential. To lowest order
$i\left[V(\bx_1)-V(\bx_2)\right]\exp[\cdots]/\hbar=\partial_\bx
V(\bx)\partial_\bp\exp[\cdots]$ so that, after taking expectation
values, one arrives at
\begin{equation}\label{Husimi3}
  \partial_tF=-\bv\cdot\partial_\bx F+\partial_\bx V\partial_\bp F+{\cal O}(\hbar)\,.
\end{equation}
To zeroth order in $\hbar$ this is the standard Liouville equation and
corresponds to equation~(\ref{eq:eom4}) for only one flavor and thus
absence of oscillations.

We note that for non-relativistic matter the Husimi transformation is
also used in the context of simulating the dynamical evolution of dark
matter distributions. There the Husimi transformation relates the Schr\"odinger-Poisson equation for a wave function in three spatial dimensions subject
to a gravitational potential to the classical description by six dimensional phase space distributions governed by Liouville equations,
also known as collisionless Vlasov equations~\cite{Skodje:1989,Widrow:1993qq,Mocz:2018ium}. Their momentum integrated
version leads to the classical equations of hydrodynamics, i.e., the
continuity and Euler equations, which in turn are related to the
Schr\"odinger equation by a so-called Madelung transformation~\cite{Madelung}. In these
contexts, classical behavior emerges on length scales large compared to the de Broglie wavelength $\hbar/p$.
The use of Schr\"odinger-like equations instead of classical equations of motion can have practical advantages in dark matter simulations:
One has to deal only with three space coordinates instead of six phase-space coordinates or a large number of
particles in $N-$body simulations. Furthermore, singularities that can develop in solutions of the classical Liouville or hydrodynamics equations,
for example at caustics and shocks, are smoothed out by a finite de Broglie wavelength.

\end{document}